\documentclass[10pt,article,aps,prd,reprint,nofootinbib,floats,floatfix,amsfonts,amssymb,amsmath,preprintnumbers,notitlepage,superscriptaddress]{revtex4-1}

\usepackage{graphicx,xcolor}
\usepackage{hyphenat} 
\usepackage{amsmath, amssymb, amsfonts,amsbsy,amsthm} 
\usepackage{float}
\usepackage{hyperref}

\newcommand{\hatbf}[1]{\mathbf{\hat{#1}}}

\newcommand{\orcid}[1]{\href{https://orcid.org/#1}{\includegraphics[scale=0.15]{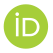}}}

\begin{document}
\title{Gravitational-Wave Geodesy:\\ Defining False Alarm Probabilities with Respect to Correlated Noise}
\author{Kamiel Janssens\orcid{0000-0001-8760-4429}}
\affiliation{Universiteit Antwerpen, Prinsstraat 13, 2000 Antwerpen, Belgium}
\affiliation{Artemis, Universit\'e C$\hat{o}$te d’Azur, Observatoire C$\hat{o}$te d’Azur, Nice, France}

\author{Thomas A. Callister\orcid{ 0000-0001-9892-177X}}
\affiliation{Center for Computational Astrophysics, Flatiron Institute, New York, NY 10010, USA}

\author{Nelson Christensen\orcid{0000-0002-6870-4202}}
\affiliation{Artemis, Universit\'e C$\hat{o}$te d’Azur, Observatoire C$\hat{o}$te d’Azur, Nice, France}
\author{Michael W. Coughlin \orcid{0000-0002-8262-2924}}
\affiliation{School of Physics and Astronomy, University of Minnesota, Minneapolis, Minnesota 55455, USA}

\author{Ioannis Michaloliakos\orcid{0000-0003-2980-358X}}
\affiliation{University of Florida, Gainesville, Florida 32611, USA}

\author{Jishnu Suresh\orcid{0000-0003-2389-6666}}
\affiliation{Institute for Cosmic Ray Research, The University of Tokyo, Kashiwa City, Chiba 277-8582, Japan}
\affiliation{Universit\'e catholique de Louvain, B-1348 Louvain-la-Neuve, Belgium}

\author{Nick van Remortel\orcid{0000-0003-4180-8199}}
\affiliation{Universiteit Antwerpen, Prinsstraat 13, 2000 Antwerpen, Belgium}

\date{\today}

\begin{abstract}

Future searches for a gravitational-wave background using Earth-based gravitational-wave detectors might be impacted by correlated noise sources. A well known example are the Schumann resonances, which are extensively studied in the context of searches for a gravitational-wave background. Earlier work has shown that a technique termed ``gravitational-wave geodesy'' can be used to generically differentiate observations of a gravitational-wave background from signals due to correlated terrestrial effects, requiring true observations to be consistent with the known geometry of our detector network. 
The key result of this test is a Bayes factor between the hypotheses that a candidate signal is astrophysical or terrestrial in origin.
Here, we further formalize the geodesy test, mapping distributions of false-alarm and false-acceptance probabilities to quantify the degree with which a given Bayes factor will boost or diminish our confidence in an apparent detection of the gravitational-wave background.
To define the false alarm probability of a given Bayes factor, we must have knowledge of our null hypothesis: the space of all possible correlated terrestrial signals. Since we do not have this knowledge we instead construct a generic space of smooth functions in the frequency domain using Gaussian processes, which we tailor to be conservative. This enables us to use draws from our Gaussian processes as a proxy for all possible non-astrophysical signals.
As a demonstration, we apply the tool to the SNR = 1.25 excess observed for a 2/3-power law by the LIGO and Virgo collaborations during their second observing run.

\vspace{1cm}
\end{abstract}

\maketitle

\section{Introduction}
\label{sec:Introduction}

Since the first detection of gravitational waves (GW) in 2015~\citep{FirstGWObs}, the Advanced LIGO~\cite{2015}, Advanced Virgo~\cite{VIRGO:2014yos} and KAGRA~\cite{PhysRevD.88.043007} collaborations (LVK collaborations) have announced many more binary mergers~\citep{GWTC1,GWTC2,NSBHDiscovery,GWTC3}. These include binary black hole mergers, binary neutron star mergers, as well as neutron star-black hole mergers. In total, 90 observations were reported by the LVK in GWTC-1~\citep{GWTC1}, GWTC-2~\citep{GWTC2}, GWTC-3~\citep{GWTC2} and some lower significant events in GWTC-2.1~\cite{GWTC2.1}.

Gravitational waves from the mergers of most binary systems at cosmological distances are too weak to be individually detected. However, the superposition of these signals forms a gravitational-wave background (GWB)~\citep{Regimbau_2008,CBC-GWB-1,CBC-GWB-2,CBC-GWB-3,CBC-GWB-4,CBC-GWB-5}.
The LVK collaboration has conducted searches for both an isotropic~\citep{O3Isotropic} and anisotropic~\citep{O3Directional} gravitational-wave background, but no such signal has yet been detected by the Advanced detectors.
Predictions based on the merger rate and mass distribution of compact binaries indicate the GWB from the superposition of these events might be detected by LIGO, Virgo and KAGRA during forthcoming observing runs~\citep{ObsProspects,O3Isotropic}. Apart from the GWB from unresolved binary mergers many other astrophysical and cosmological signals could contribute to a GWB~\citep{SGWBRevChristensen}.

As the GWB is too weak to be observed in a single detector, we rely instead on the cross-correlation of strain data from multiple detectors~\citep{SGWBRevRomano,SGWBRevChristensen}. If we assume the noise present at different detector sites is uncorrelated, any excess correlation between the strain measured in two detectors must be due to an astrophysical gravitational-wave signal. It is not, however, the case that gravitational-wave detectors measure strictly independent noise realizations. While many sources of terrestrial noise are indeed local, there are known sources that are correlated on global scales, introducing non-astrophysical correlations in the LIGO-Hanford, LIGO-Livingston, Virgo, and KAGRA interferometers, despite their separation by several thousand kilometers.
Some known examples of such noise sources are Schumann resonances~\citep{Schumann1,Schumann2} and the synchronization of on-site electronics to  Global Positioning System (GPS) time~\citep{Covas2018}.
Schumann resonances are electromagnetic excitations in the cavity formed by the Earths-surface and the ionosphere, sourced by lightning strikes across the globe~\citep{Schumann1,Schumann2}. They are expected to couple magnetically to the interferometers and induce a correlated signal of terrestrial origin~\citep{Thrane2013,Thrane2014}.

The existence of known sources of correlated noise raise an important question: If we detect evidence for a correlated signal between two (or more) GW interferometers, how can we be confident that this source is of astrophysical origin rather than terrestrial?
Until now, several methods have been or are being developed to help differentiate between a correlated signal due to gravitational waves or due to terrestrial sources. For Schumann resonances, in particular, methods have been investigated to directly measure and remove their effect by applying Wiener filtering~\citep{Thrane2013,Thrane2014,Coughlin2016,Coughlin2018} and more recently to incorporate Schumann resonances and a GW signal in one consistent Bayesian parameter estimation framework~\citep{BayesianGW-Mag}.

A complementary method, gravitational-wave geodesy (GW-Geodesy), was previously proposed~\citep{GeodesyOriginal}.
In GW-Geodesy, the geometry of an interferometer network (the relative distances and orientations of component detectors) is reversed-engineered from an observed GWB. This forms the basis for a consistency check that a true astrophysical signal must pass. A true GW signal must yield results consistent with the \textit{known} geometry or our baseline, while the same is not the case for other sources. In the first implementation of the GW-Geodesy framework, it was shown that the method can successfully differentiate an isotropic GWB coming from unresolved binary mergers from correlated terrestrial noise due to Schumann resonances or the synchronization of electronics to GPS-time~\citep{GeodesyOriginal}.

The output of the GW-Geodesy test is a Bayes factor between two hypotheses: (\textit{i}) that a tentative detection yields consistency with our known baseline geometry and is therefore astrophysical, or (\textit{ii}) that the signal prefers an unphysical geometry and is hence non-astrophysical in origin.
This Bayes factor acts as a secondary test statistic, complementary to the signal-to-noise ratio (SNR) with which we observe a given signal.
While these Bayes factors can be qualitatively interpreted, until now we have not been able to assign a precise \textit{statistical significance} to a given Bayes factor.
Ideally, we could assign any given Bayes factor a false alarm probability (FAP) and a false dismissal probability (FDP).
The FAP indicates how often one might accidentally confirm a terrestrial signal based on this test, whereas the FDP indicates how often we accidentally dismiss a real signal.
In this work, we quantity these FAP and FDPs, allowing us to answer the crucial question: \textit{How likely is an observed correlated  signal with a given SNR and geodesy Bayes factor to be of astrophysical origin, rather than a yet-unidentified source of terrestrial correlation?}
In this fashion, we can utilize GW-Geodesy not only as a tool with which to reject terrestrial signals, but also as one to bolster our confidence in a real gravitational-wave background detection.

To be able to construct FAPs and FDPs, we first need a proxy for unknown correlated terrestrial signals, which have the possibility of contaminating the isotropic stochastic search.
This is, by definition, challenging: we cannot know the nature of unknown contaminants.
We therefore instead utilize Gaussian processes to represent random and a priori unknown contaminants, defining FAPs and FDPs over the space of continuous cross-correlation functions that LVK detectors might measure.

In Sec.~\ref{sec:Math}, we will cover some mathematical concepts of stochastic searches and the GW-Geodesy framework which are crucial elements in this work. In Sec.~\ref{sec:GP}, we introduce Gaussian processes, and we optimize the model parameters to create a worst case scenario. In Sec.~\ref{sec:demonstration}, we will demonstrate how the Gaussian processes can be used within the framework of GW-Geodesy and why it can become a powerful tool to help in validating future observations of an isotropic stochastic background. As part of this demonstration we apply our framework in Sec.~\ref{sec:O2outlier} to the SNR = 1.25 excess observed for a 2/3-power law by the LIGO and Virgo collaborations during their second observing run (O2). In Sec.~\ref{sec:DiscussionOutlook}, we discuss the tool together with an outlook on possible future additions or improvements.

\section{Gravitational-Wave Geodesy}
\label{sec:Math}

We often characterize the GWB in terms of its energy-density spectrum $\Omega (f)$ -- Eq. \ref{eq:energyDensity} -- expressed as the energy density $d\rho_{GW}$ of GWs per logarithmic frequency interval $d \ln(f)$. $\Omega (f)$ is made dimensionless by dividing by the Universe's critical energy density $\rho_c = 3H_0^2c^2/(8\pi G)$, where $H_0$ is the Hubble constant, $c$ the speed of light, and $G$ is Newton's constant \citep{SGWBRevAllen,SGWBRevRomano}:
\begin{equation}
\label{eq:energyDensity}
\Omega(f) = \frac{1}{\rho_c}\frac{d\rho_{GW}}{d \ln(f)}
\end{equation}

To measure the energy density of the GWB, one computes the cross-correlation spectrum $\hat{C}(f)$ between two gravitational-wave observatories. If $\Tilde{s}_I(f)$ is the measured (Fourier domain) strain of observatory $I$ and $\Delta T$ the duration of the analyzed data, one can express  $\hat{C}(f)$ as 
\begin{equation}
\label{eq:CrossCorr}
\hat{C}(f) = \frac{1}{\Delta T}\frac{20\pi^2}{3H_0^2}\mathrm{Re}[\Tilde{s}_1^*(f)\Tilde{s}_2(f)].
\end{equation} 
The normalisation of $\hat{C}(f)$ is chosen such that its expectation value is given by \citep{SGWBRevAllen}
 \begin{equation}
\label{eq:ExpectationValue}
\langle \hat{C}(f) \rangle = \gamma(f) \Omega(f).
 \end{equation}
 $\gamma(f)$ is the normalized overlap reduction function, which encodes the imprint of the detectors' baseline geometry (location and relative orientations) on the observed correlations. For two laser interferometers, the normalized overlap reduction is given by \citep{SGWBRevChristensen1992}
 \begin{equation}
 \label{eq:ORF}
 \gamma(f) = \frac{5}{8\pi}\sum_A \int_{\rm Sky} F^A_1(\hatbf{n})F^A_2(\hatbf{n})e^{2\pi if\Delta\textbf{x}\cdot\hatbf{n}/c}d\hatbf{n}.
 \end{equation}
  $F^A_I(\hatbf{n})$ is the antenna response of detector I to GWs with polarization A. $\Delta x$ represents the separation vector between the two detectors, whereas $\hatbf{n}$ indicates the sky direction. One sums over all tensor polarizations (``plus'' and ``cross'') and integrates over all sky-directions. By virtue of the leading factor $\frac{5}{8\pi}$, co-located and co-aligned detectors will have $\gamma(f)=1$ for all frequencies.
  
  If we assume that we are in the weak signal limit (which is a valid assumption since the GWB has yet to be observed), the co-variance of $\hat{C}(f)$ at two different frequencies $f$ and $f'$ is given by $\langle \hat{C}(f) \hat{C}(f') \rangle=\delta(f-f')\sigma^2(f)$ where $\sigma^2(f)$ is given by \citep{SGWBRevAllen,SGWBRevRomano}
  \begin{equation}
  \label{eq:Variance}
  \sigma^2(f) = \frac{1}{\Delta T}     \left( \frac{10\pi^2}{3H_0^2}\right)^2 f^6P_1(f)P_2(f).
  \end{equation}
  and $P_I(f)$ is the noise power spectral density of detector I.
  
 Traditionally, searches for a GWB assume a power-law of the form:
  \begin{equation}
     \Omega(f)=\Omega_{\rm ref}\left(\frac{f}{f_\mathrm{ref}}\right)^{\alpha}.
 \end{equation}
  A power-law index $\alpha = 0$ is expected from several cosmological sources of a GWB, while $\alpha = 2/3$ is expected from individually unresolved binary coalescence events. A GWB with $\alpha = 3$ could be produced by supernovae~\cite{SGWBRevChristensen,SGWBTania}.
  
Given the detection of a gravitational-wave background, we could seek to infer $\Omega_{\rm ref}$ and $\alpha$. Additionally, however, the dependence of our measured cross-correlation spectrum $\hat{C}(f)$ on the overlap reduction function means that the GWB could be used to infer the geometry of our detector network itself. This fact forms the basis of GW geodesy: a true GWB should yield an inferred geometry consistent with the true geometry of our detector network. An isotropic astrophysical/cosmological GWB must be consistent with the expected functional form of our baseline's overlap reduction function. Correlated terrestrial noise sources, on the other hand, do not necessarily need to follow the behaviour of the overlap reduction function. Thus, there is no reason why non-GW correlated noise sources would prefer the true geometry over any random geometry  \citep{GeodesyOriginal}.
 
We formalize this test by defining the following two hypotheses:

 \begin{itemize}
     \item Hypothesis \( \mathcal{H}_{\gamma} \): The measured cross-correlation is consistent with the \textit{true} baseline geometry and overlap reduction function.
     \item Hypothesis \( \mathcal{H}_{\mathrm{Free}} \): The measured cross-correlation spectrum is consistent with an \textit{unphysical} baseline geometry. Under this hypothesis, we treat our detector positions and orientations as free variables to be inferred from the data, allowing them to range (unphysically) across the surface of the Earth.
 \end{itemize}
 
To compare the hypotheses $\mathcal{H}_{\gamma}$ and $\mathcal{H}_{\mathrm{Free}}$, one can construct a Bayes factor $\mathcal{B}$ between these hypotheses to establish which model is favored by the cross-correlated data $\hat{C}$,
 \begin{equation}
 \label{eq:logBayes}
     \mathcal{B} = \frac{p(\hat{C}|\mathcal{H}_{\mathrm{\gamma}})}{p(\hat{C}|\mathcal{H}_{\mathrm{\mathrm{Free}}})},
 \end{equation} 
 where $p(\hat{C}|\mathcal{H}_{\mathrm{\gamma}})$ and $p(\hat{C}|\mathcal{H}_{\mathrm{\mathrm{Free}}})$ are the probabilities of finding the observed cross-correlation -- as defined in Eq. \ref{eq:CrossCorr} -- given hypothesis $\mathcal{H}_{\mathrm{\gamma}}$ and $\mathcal{H}_{\mathrm{\mathrm{Free}}}$, respectively.
  Because $\mathcal{H}_{\mathrm{\mathrm{Free}}}$ is a more complex model, it will be penalized by the Bayesian ``Occam's factor''. Therefore, an isotropic astrophysical signal will be consistent with both  $\mathcal{H}_{\gamma}$ and  $\mathcal{H}_{\mathrm{Free}}$, but will favor  $\mathcal{H}_{\gamma}$ since it is the simpler hypothesis. Non-GW correlated noise sources have a priori no preference for  $\mathcal{H}_{\gamma}$ and therefore will be better fit by the additional degrees of freedom provided by non-physical geometries, leading to favouring the  $\mathcal{H}_{\mathrm{Free}}$ hypothesis.

\section{Gaussian processes}
\label{sec:GP}

If, in the future, we compute a geodesy Bayes factor $\mathcal{B}$ associated with a candidate gravitational-wave background, we do not yet know exactly how we should quantitatively interpret this result. In order to fully understand the statistical significance of a particular Bayes factor, we need to know how often it arises simply by chance. This poses a dilemma, however.
To quantify a false-alarm probability, we would need to know all possible terrestrial signals, which we do not. So instead, we can ask a similar question: how often do particular Bayes factors arise from the generic space of smooth functions in the frequency domain? This latter question can be answered using Gaussian processes.

Gaussian processes are very flexible and are frequently used for model fitting or model predictions based on a data set \cite{GPForMachineLearning}. In this work, we use the Gaussian processes to produce a distribution of functions with a given mean and variance. We consider each draw from this distribution to be a possible realisation of a correlated terrestrial signal. 

The two main inputs of the Gaussian process are a co-variance matrix, often called the \textit{kernel}, and the mean. In this work, we use the Gaussian processes to generate possible realizations of $\langle \hat{C}(f) \rangle$. The probability density function for the draws is given in Eq.~\ref{eq:pdf}. This is the probability to draw a certain cross-correlation spectrum \textbf{C(f)}, from the function space governed by the co-variance matrix $\Sigma$ \citep{GPForMachineLearning}
\begin{equation}
\begin{aligned}
\label{eq:pdf}
    p(\textbf{C(f)}|\mu,\Sigma) = &\frac{1}{(2\pi)^{n/2}|\Sigma|^{1/2}} \cdot \\
                            & \mathrm{exp}\left( - \frac{1}{2}(\textbf{C(f)}-\mu)^T \Sigma^{-1} (\textbf{C(f)}-\mu)\right)
\end{aligned}
\end{equation}
In Eq.~\ref{eq:pdf}, $n$ is the dimension of the co-variance matrix $\Sigma$. 

We generate distributions of cross-correlation spectra with zero mean and consider one of the most commonly used co-variance matrices, the so called \textit{squared exponential (SE)}. In this case, the covariance between cross-correlation values measured at frequencies $f_i$ and $f_j$ is given by~\citep{GPForMachineLearning}:\\
\begin{equation}
\label{eq:basiskernels}
    \Sigma_{\mathrm{SE},ij}(\sigma,l) = \sigma^2 \mathrm{exp}\left( - \frac{(f_i-f_j)^2}{2l^2} \right). 
\end{equation}
$\sigma^2$ is the variance at a single frequency, and while this parameter can be used to scale the signal strength with respect to observations, it will have no impact on the spectral \textit{shape} of the draws from the distribution.
$l$ is the characteristic length-scale over which our \textbf{C(f)} measurements are correlated.

In order to utilize Eqs. 8 and 9, we still need to choose a length-scale parameter $l$. We will tune this parameter by deliberately targeting the 'worst-case scenario', in which the Gaussian process yields cross-correlation spectra that look -- on average -- most like a proper astrophysical/cosmological signal. In doing so, we will always obtain conservative estimates of the false-alarm probability for a given geodesy Bayes factor.

To select the 'worst-case' value of the length-scale parameter, we will maximize the probability for our \textit{target astrophysical signal} itself to be drawn from the Gaussian process. The cross-correlation spectrum we expect from a gravitational-wave background is $\textbf{C(f)}=\gamma_{HL}\cdot \Omega_{ref}\left(\frac{f}{f_{ref}}\right)^{\alpha} :=  \bf{C_{\alpha}}\mathrm{(f)}$, with $\alpha =$ 0, 2/3 or 3, where we approximate the mean to be zero. These are the power-laws typically looked for by the LIGO, Virgo and KAGRA collaborations \citep{O3Isotropic}. Note that we have chosen the Hanford-Livingston (HL) baseline as our observing baseline. This baseline is the most sensitive for observing an isotropic GWB, due to their better sensitivity compared to Virgo but more importantly because their overlap reduction function is considerably larger.
The log-likelihood of this signal under our Gaussian process is
\begin{equation}
\label{eq:MaxLikelihood}
\begin{aligned}
        \ln(\mathcal{L}) =& \ln p(\bf{C_{\alpha}}\mathrm{(f)}|\mu=0,\Sigma) \\ 
        =& -\frac{1}{2} \ln(|\Sigma|)  - \frac{1}{2} \bf{C_{\alpha}}\mathrm{(f)}^T \Sigma^{-1} \bf{C_{\alpha}}\mathrm{(f)} + \mathrm{constants}.
\end{aligned}
\end{equation}

As we will later freely adjust the overall amplitude of our Gaussian process draws in order to vary their SNRs, we will optimize only the length parameter $l$, fixing the overall covariance to $\sigma^2 = 1$.
Accordingly, for consistency we normalize our target astrophysical signal via
    \begin{equation}
    \tilde{\bf C}_\alpha(f) = \frac{\bf C_\alpha(f)}
        {\sqrt{\mathrm{Var}\,\bf C_\alpha(f)}},
    \end{equation}
choosing the $l$ that maximizes $\ln p( \tilde{\bf C}_\alpha(f)|\mu=0,\Sigma)$.

In practice, the inversion of $\Sigma$ is unstable, with a determinant that is nearly zero. 
To increase stability, we add a diagonal term to the co-variance matrix: $\Sigma \rightarrow \Sigma + \epsilon\, I$, with $I$ the identity matrix and $\epsilon$ a small dimensionless constant, here chosen to be $10^{-3}$. This diagonal term can be interpreted as a noisy observation of our function $\bf{C_{\alpha}}\mathrm{(f)}$.

We optimize the kernel parameter $l$ using the formalism described above, in the following parameter range: $l \in [ 10^{-4}\,\mathrm{Hz},10^{4}\,\mathrm{Hz}]$. 
The optimal length-scale parameter to mimic a power-law with slope $\alpha=$ 0, 2/3 and 3 are respectively: $l = 47.65\,\mathrm{Hz}$, $33.18\,\mathrm{Hz}$ and $29.20\,\mathrm{Hz}$, determined using a 0.01Hz frequency resolution for $l$.

\begin{figure}
\centering
\includegraphics[width=0.48\textwidth]{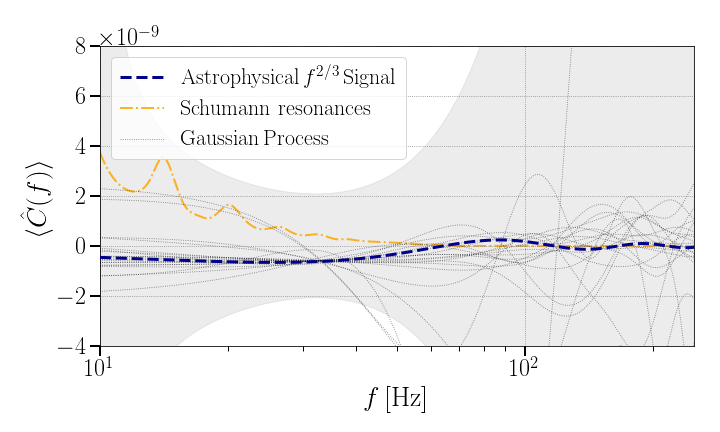}
\caption{
The expected cross-correlation for an astrophysical background from unresolved binary systems (blue/bold), Schumann resonances (orange) and 20 random draw from the $SE$ kernel (gray). The kernel-parameter used for the Gaussian process instantiations is the optimal value for mimicking an 2/3-power-law signal, as described in the text. The injected signal strength yields a $\mathrm{SNR}=3$ after a 3-year long observation at LIGO's design sensitivity.
}
\label{fig:randomdraws}
\end{figure}

In Fig.~\ref{fig:randomdraws}, we show several random draws from the Gaussian process described by the SE-kernel, optimized to most closely match an $\alpha=2/3$ astrophysical signal, as well as the signals we expect to observe from unresolved binary mergers and terrestrial Schumann resonances. Here we have chosen $\alpha=2/3$ since this gravitational-wave background is the signal expected to be first observable by advanced LIGO and advanced Virgo.

This figure illustrates the SE-kernel with optimized parameters is able to produce cross-correlation curves which are, on average, functionally similar to the true-signal curve. As discussed above, this is deliberately conservative, maximizing the false-alarm probabilities with which our Gaussian process will randomly yield cross-correlation functions with favorable geodesy Bayes factors. As already was stated in the initial implementation of the GW-Geodesy framework, having similar zero-crossings might play a crucial role in mimicking a signal \citep{GeodesyOriginal}.
Everywhere in this analysis we investigate the GW-Geodesy framework in the frequency range of 10--250\,Hz with a frequency resolution of 0.24\,Hz. The lower limit is based on the lowest frequency that can be observed by current ground-based interferometers. Because of the combined effects of the overlap reduction function and the reduced sensitivity at higher frequencies, there is a negligible gain when going to higher frequencies compared to the computational costs. For example, when looking for an isotropic power-law signal using the HL-baseline with a power-law slope of $\alpha=$0, 2/3 and 3, 99\% of the sensitivity during the latest O3 run was respectively contained below 76.1\,Hz, 90.2\,Hz and 282.8\,Hz~\citep{O3Isotropic}.

In the rest of this paper, when we mention a Gaussian process signal, we are referring to a Gaussian process with a SE kernel and a length-scale parameter optimized for the relevant power-law slope $\alpha$. Also, when mentioning ``unknown correlated noise sources,'' we refer to this Gaussian process signal, which in this study serves as a conservative proxy for these unknown correlated noise sources.

\section{False alarm probabilities and detection confidence}
\label{sec:demonstration}

\subsection{Simulations}
Using our Gaussian process machinery, we will explore the false alarm probabilities and statistical significance associated with geodesy Bayes factors. We will make use of three different sources of cross-correlation: a power-law signal with slope $\alpha$, magnetic Schumann resonances and a proxy for unknown terrestrial cross-correlation mimicking an $\alpha$-power-law signal using Gaussian processes.
As a proof of concept we will start by investigating Bayes factors given by signals with SNR=3 after three years of observation at LIGO's design sensitivity~\citep{LIGOSensitivity}. This specific signal-to-nose-ratio is chosen as this is the fiducial value when one might first claim evidence for a detected GWB.

The signal-to-noise ratio of our model signal $\bf{C_{\alpha}}(f)$ is given by
\begin{equation}
\label{eq:SNR}
    \begin{aligned}
    \mathrm{SNR}_{\alpha} &= \frac{\int_f \hat{C}(f)\bf{C_{\alpha}}(f)\times \sigma^{2}_{\alpha}/ \sigma^{2}}{ \sigma_{\alpha}},\\
    &\mathrm{with } \\
    \sigma^{2}_{\alpha} &= \left( \int_f{\frac{\bf{C_{\alpha}}^2(f)}{\sigma^2}} \right)^{-1}.\\
    \end{aligned}
\end{equation}
In what follows we will drop the subscript $\alpha$ and refer to the signal to noise ratio SNR$_\alpha$ just as SNR. One can compute the \textit{expected} SNR or the \textit{observed} SNR. In the former case $\hat{C}(f)$ is the injected signal, whereas in the latter case Gaussian-distributed noise consistent with Eq. \ref{eq:Variance} has been added to each frequency bin of $\hat{C}(f)$.\\
When performing an injection for a certain power-law with slope $\alpha=0,2/3,3$, the signal is injected with an expected SNR$_{\alpha}$ of the desired strength. Note that in case of the Schumann signal as well as the Gaussian process signal, sometimes a power-law with a different slope might be recovered with a higher SNR: $\mathrm{SNR}_{\tilde{\alpha}}>\mathrm{SNR}_{{\alpha}}$, where $\tilde{\alpha}\neq \alpha$.

Note that the SNR defined in Eq. \ref{eq:SNR} can be positive (if the data well-match our signal model) or negative (if data anticorrelate with our signal model). As we are concerned primarily with signals we might \textit{mistake} as astrophysical, we will only investigate those that yield positive SNRs. Therefore, if the SNR of a Gaussian process draw is negative at the target $\alpha$ (0, 2/3 or 3), it is rejected and a new signal is simulated. This happens, on average, 50\% of the time. The rejection of this signal is chosen to match a realistic experimental condition. In case of a detection, one would like to apply this tool mainly to a positive SNR detection, whereas a negative SNR detection would immediately be categorized as unphysical.

\begin{figure}
\centering
\includegraphics[width=0.48\textwidth]{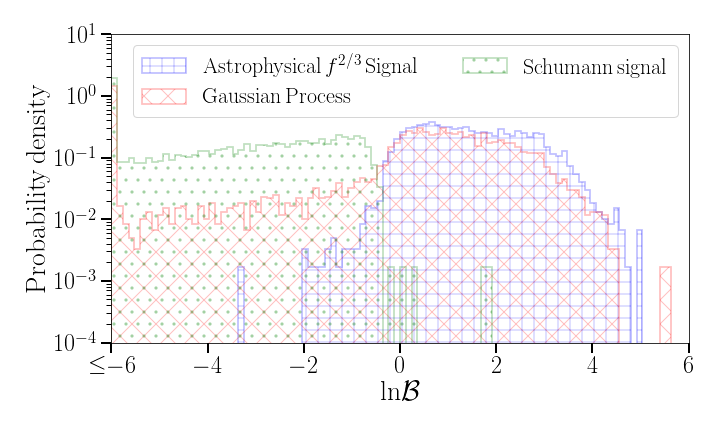}
\caption{Probability density of ln$\mathcal{B}$ for an astrophysical $\alpha=2/3$ power-law signal, random draws from a Gaussian process using our optimized (worst-case) SE-kernel and a Schumann signal. The lowest bin contains all ln$\mathcal{B} \leq -6$. For each signal type 5000 injections were performed with an injection strength of SNR=3, when recovered with an $\alpha=2/3$ signal model.}
\label{fig:SNR3_Hist}
\end{figure}

For our three signal classes, we simulate and analyze 5000 simulated cross-correlation spectra. In the case of the astrophysical power law and Schumann resonances, this involves generating 5000 distinct Gaussian noise realizations that are added to the fixed underlying models. Under our Gaussian process, meanwhile, each trial involves a random draw from our Gaussian process (restricted to positive SNR) and a randomly generated noise spectrum. For every injection, we use \texttt{PyMultiNest} to compute Bayesian evidences~\cite{refId0}, using 2000 live-points. \texttt{PyMultiNest} is a python interface for \texttt{MultiNest}~\cite{10.1111/j.1365-2966.2007.12353.x,10.1111/j.1365-2966.2009.14548.x}, which is an implementation of the nested sampling algorithm~\cite{10.1214/06-BA127,doi:10.1063/1.1835238}.

The model corresponding to our hypothesis \( \mathcal{H}_{\gamma} \) has two free parameters, the reference amplitude $\Omega_{\mathrm{ref}}$ of the signal at $f_{\mathrm{ref}}=$ 25Hz and the power-law slope $\alpha$. We use a log uniform prior for the reference amplitude between $10^{-12}$ and $10^{-6}$. For $\alpha$ we use a Gaussian prior with a standard deviation of 3.5.
Our alternative hypothesis \( \mathcal{H}_{\mathrm{Free}} \) has three additional free parameters, the distance between the two interferometers $\Delta x$ and the rotation angles of your interferometers $\phi_1$ and $\phi_2$~\cite{GeodesyOriginal}. We use uniform priors on $\phi_1$ and $\phi_2$ (0,2$\pi$). Furthermore we use a uniform prior on cos$\theta$, where $\Delta x = 2R_{\mathrm{Earth}}sin\theta/2$. This corresponds to a prior on the distance between the detectors: p($\Delta x$)$\propto \Delta x$. These priors are chosen to be consistent with earlier work~\citep{GeodesyOriginal}.

Figure \ref{fig:SNR3_Hist} represents the log-Bayes distribution for the different signal models, assuming a SNR=3 and $\alpha=2/3$. 
The lower bin in the histograms also includes all simulations with a recovered log-Bayes factor smaller than -6. The smallest log-Bayes factor for an injection with a Schumann signal is approximately -23, whereas this is approximately $-6\times 10^7$ in case of the Gaussian process signal.

First, we notice the overlap of the histograms of the Gaussian process and the power-law signals is significantly larger than the overlap between the Schumann signal and the power-law signal. This shows the selection of the Gaussian process parameters in section \ref{sec:GP} is successful and indeed yields a conservative condition, where the Gaussian process is able to mimic the power-law signal. However, it also shows what possibly is one of the weaknesses of current implementation of the tool: our Gaussian process might be overly conservative as it is able to very well mimic the $\alpha = 2/3$-power-law, and therefore to a large extent yielding similar log-Bayes factors.

On the other hand, whereas power-law signals yield a handful of mildly negative log-Bayes values, the Gaussian processes give an extended tail towards negative log-Bayes factors, which will further grow as we increase the SNR of our injections. This illustrates the intrinsic random nature of the Gaussian process. Despite the process being able to produce signals mimicking the power-law signal, at the same time, other types of signals are produced which are not properly described by a power-law.

Assuming the Gaussian process is a (very) conservative, estimate of a terrestrial contamination for isotropic GWB searches, we can construct an upper bound on the false alarm probability (FAP) and detection probability associated with our Geodesy Bayes factors.
The FAP is the probability with which our terrestrial signal (for which the Gaussian process is our proxy) gives a Geodesy Bayes factor as high or higher than the Bayes factor we recover from our actual data. The detection probability gives the probability to detect the signal at a certain FAP/log-Bayes factors. The detection probability and False Dismissal Probability (FDP) are linked to each other by: $\mathrm{FDP} = 1 - \mathrm{det. prob.}$; where the FDP is a figure of merit of the probability to wrongly reject the signal model.

For future observations, we would like to have an as small as possible FAP (unlikely to be a false signal) and a large detection probability (likely to be true signal). Generally, before analysing the data, one chooses a FAP considered to be the largest allowed value, for example 5\% or 1\%. Given a signal injected with a SNR=3 the log-Bayes factors and detection probability are shown in Table \ref{tab:FAP/FDP} for a FAP of 5\% and 1\%. We also show the results when looking for a power-law signal with slope $\alpha = $ 0 or 3.
We see, for example, that given an apparent detection of the gravitational-wave background with $\mathrm{SNR}=3$ under an $\alpha=2/3$ model, there is no more than a 5\% chance that a log-Bayes factor $\ln\mathcal{B}=2.94$ would arise by chance from a non-astrophysical signal.

We notice that for a given FAP, the detection probability becomes higher for power-law signals with a steeper slope (larger $\alpha$). However all reported detection probabilities are very small. This is linked to our Gaussian process being (overly) conservative and is very good in mimicking the GW power-law signals. This was already clear from Figure \ref{fig:SNR3_Hist} for $\alpha=2/3$. 

Table \ref{tab:FAP/FDP-SNR5} shows the same for an injected signal with SNR=5 when looking for a power-law with slope $\alpha=$ 2/3 or 3. The detection probability for a 2/3-power-law remains very small, for an $\alpha =3$ power-law there is on the other hand a drastic increase in detection probability when going from SNR=3 to SNR=5.

Table \ref{tab:FAP/FDP_Schumann}, for comparison, shows log-Bayes factors and detection probabilities when one compares the power-law signals with a Schumann signal. For a fixed FAP of 1\%, the detection probability is $>$99\% meaning the GW-Geodesy tool is very effective in differentiating a Schumann signal from a power-law signal at a SNR of 3.

\begin{table}
    \centering
    \begin{tabular}{|c|c|c|c|c|c|c|}
    \cline{2-7}
       \multicolumn{1}{c|}{} & \multicolumn{2}{|c|}{$\alpha = 0$} & \multicolumn{2}{|c|}{$\alpha = 2/3$} & \multicolumn{2}{|c|}{$\alpha = 3$} \\
       \cline{2-7}
       \hline
        FAP  & ln$\mathcal{B}$ & det. prob. & ln$\mathcal{B}$ & det. prob. & ln$\mathcal{B}$ & det. prob.      \\
       \hline
       \hline
        5.0\% &  2.95  & 6.92\% & 2.94 & 10.56\% & 3.42  & 15.44\% \\ 
        1.0\% &  3.68  & 0.76\% & 3.72 & 1.78\% & 4.04  & 8.10\%\\
        \hline
    \end{tabular}
    \caption{The log-Bayes factor and detection probabilities matching a FAP of 1\% and 5\% comparing a $\alpha$-power-law GW signal and a Gaussian process signal. For each signal type 5000 injections were performed with an injection strength of SNR=3.}
    \label{tab:FAP/FDP}
\end{table}

\begin{table}
    \centering
    \begin{tabular}{|c|c|c|c|c|}
    \cline{2-5}
       \multicolumn{1}{c|}{} & \multicolumn{2}{|c|}{$\alpha = 2/3$} & \multicolumn{2}{|c|}{$\alpha = 3$} \\
       \cline{2-5}
       \hline
        FAP  & ln$\mathcal{B}$ & det. prob. & ln$\mathcal{B}$ & det. prob.       \\
       \hline
       \hline
        5.0\% &  3.67 & 15.54\% & 3.91  & 67.08\% \\ 
        1.0\% &  4.30  & 1.80\% & 4.71 & 41.40\%\\
        \hline
    \end{tabular}    
    \caption{The log-Bayes factor and detection probabilities matching a FAP of 1\% and 5\% comparing a $\alpha$-power-law GW signal and a Gaussian process signal. For each signal type 5000 injections were performed with an injection strength of SNR=5.}
    \label{tab:FAP/FDP-SNR5}
\end{table}

\begin{table}
    \centering
    \begin{tabular}{|c|c|c|c|c|}
    \cline{3-5}
       \multicolumn{2}{c|}{} & $\alpha = 0$ & $\alpha = 2/3$ & $\alpha = 3$ \\
       \cline{3-5}
       \hline
       ln$\mathcal{B}$ & FAP  & \multicolumn{3}{|c|}{det. prob.}     \\
       \hline
       \hline
       
       -0.82  & 5.0\% & 99.72\% & 99.64\% &  99.60\% \\ 
       -0.55  & 1.0\% & 99.44\%  & 99.28\% &  99.28\% \\
        \hline
    \end{tabular}
    \caption{The log-Bayes factor and detection probabilities matching a FAP of 1\% and 5\% comparing a $\alpha$-power-law GW signal and a Schumann signal. For each signal type 5000 injections were performed with an injection strength of SNR=3.}
    \label{tab:FAP/FDP_Schumann}
\end{table}

As searches for the gravitational-wave background accumulate SNR slowly over the course of months to years, the above scenario in which a candidate signal has moderate $\mathrm{SNR}=3$ represents a realistic situation in which we will first need to use the Geodesy test.
It is instructive, however, to more broadly investigate how the distributions of log-Bayes factors evolve as a function of SNR.
To do this, we simulate signals with strengths logarithmically spaced between SNR=0.1 and SNR=100. Here we will only be looking at the $\alpha = 2/3$ case, injecting astrophysical $\alpha=2/3$ power laws and drawing random ``terrestrial'' signals from our Gaussian process optimized to this same power-law form. 

As mentioned above, the free parameters are the reference signal strength $\Omega_{\mathrm{ref}}$ at 25 Hz and the power-law slope $\alpha$ in case of our hypothesis \(\mathcal{H}_{\gamma} \). For the hypothesis \(\mathcal{H}_{\mathrm{Free}} \), the set of five free parameters consist of $\Omega_{\mathrm{ref}}$, $\alpha$, $\Delta x$, $\phi_1$ and $\phi_2$. Note that for our astrophysical signal, even though we inject $\alpha$=2/3 our best fit $\alpha$ may well differ due to different noise instantiations, whereas for the Gaussian process realizations alpha will adjust to best fit the random signal with a power law.

\begin{figure}
\centering
\includegraphics[width=0.48\textwidth]{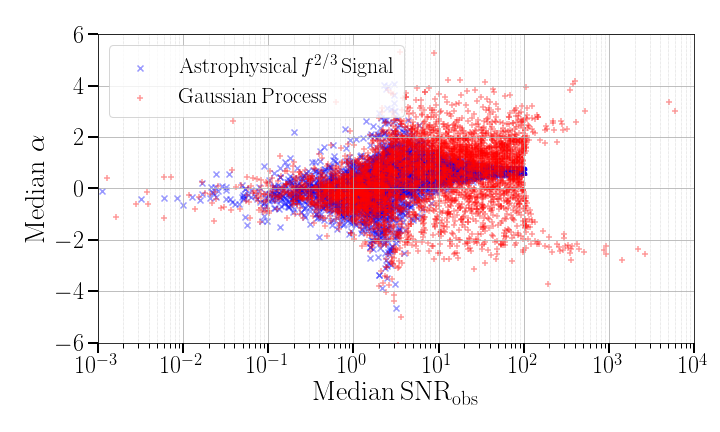}
\caption{The median of the posterior of the power-law slope $\alpha$ is represented against the median of the SNR-posterior, assuming the signal has been observed by the HL-baseline. Each point in this scatter plot represents one injected signal. For each signal type 5000 injections were performed with a logarithmically spaced injection strength between SNR=0.1 and SNR=100.}
\label{fig:SNRVar_MedianAlpha}
\end{figure}

In Figure \ref{fig:SNRVar_MedianAlpha}, the median of the $\alpha$-posterior is shown with respect to the SNR at which this signal would be observed by the HL-baseline. Note that both $\alpha$ and SNR are calculated from the posterior consistent with the HL-baseline. Although this might be disfavoured with respect to the posteriors from the Random-Baseline, we are interested on how the HL-Baseline would observe such a signal.

For extremely small SNR ($<1$), $\alpha$ is closely centered around 0 for both the true power-law GW signal as well as the simulated Gaussian process signal. These signals are so weak that whatever we observe is dominated by the Gaussian background of our search. With respect to itself this Gaussian background has by definition no power-law slope and therefore matches $\alpha = 0$.

When our signal has a strength of order SNR 1, the retrieved values of $\alpha$ are still centered around zero, but the variance increases. The posterior-median $\alpha$ spans a range from large negative to large positive power-law slopes. There starts to be some excess but given the weak strength of the signal the randomness of the Gaussian background can drive the large negative or positive $\alpha$.

The behaviour for power-law GW signals and the Gaussian process signals starts to differ from SNR $\sim$ 10. In the case of a power-law signal, the variation on $\alpha$ drastically decreases and the center of the retrieved values shifts from 0 to the real value: $2/3$. This kind of behaviour is not present in case of the Gaussian process signal, which remains centered around zero with a large variation.

This seems to indicate that with large enough SNR, we can make a clear distinction between a power-law GW signal and some unknown (terrestrial) correlated noise sources, simulated by the Gaussian processes.

\begin{figure}
\centering
\includegraphics[width=0.48\textwidth]{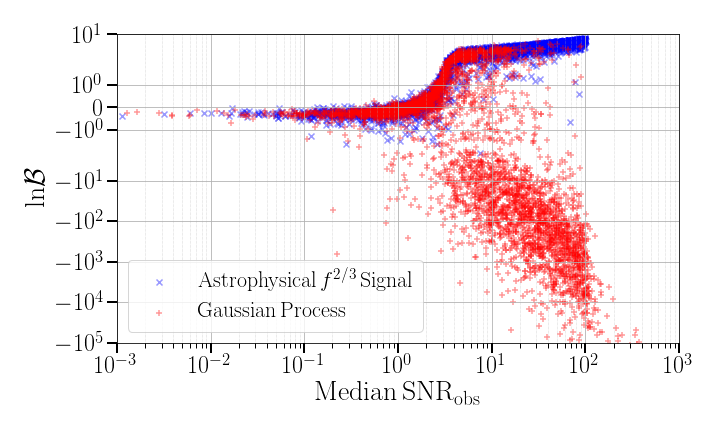}
\caption{The Log-Bayes factor comparing a 2/3-power-law signal with a Gaussian process is represented against the median of the SNR-posterior, assuming the signal has been observed by the HL-baseline. There are 40 events with ln$\mathcal{B} < 10^{-5}$, which are not shown in this figure. The smallest ln$\mathcal{B}= -5.3 \times 10^7$. Each point in this scatter plot represents one injected signal. For each signal type 5000 injections were performed with a logarithmically spaced injection strength between SNR=0.1 and SNR=100. }
\label{fig:SNRVar_LogBayes}
\end{figure}

To further strengthen this statement, Figure \ref{fig:SNRVar_LogBayes} shows the log-Bayes factor between \(\mathcal{H}_{\gamma} \) and \(\mathcal{H}_{\mathrm{Free}}\) as a  function of the observed SNR. Weak signals are on average unable to differentiate between \(\mathcal{H}_{\gamma}\) and \(\mathcal{H}_{\mathrm{Free}}\), leading to log-Bayes factors $\sim$0. When the observed SNR reaches values $\sim$1, the log-Bayes factor of the power-law $2/3$ signals starts to prefer positive values, with increasing LogBayes for increasing SNR. This is consistent with both our expectations for a true signal as well as the earlier results of the GW-Geodesy tool \citep{GeodesyOriginal}.

At the same time, the data from the Gaussian process starts to separate into two categories when the SNR reaches order one. One population of signals starts to prefer negative log-Bayes factors, preferring \(\mathcal{H}_{\mathrm{Free}}\) over \(\mathcal{H}_{\gamma}\). These are the signals that look nothing like a power-law signal as observed by the HL-baseline. However, there is also a population of signals that starts to prefer positive log-Bayes factors. These are the signals that succeeded in mimicking the HL-baseline $2/3$ power-law signal to a (very) good extent. This behaviour is expected because we purposely chose our Gaussian process parameters to have this kind of behaviour, which is the key ingredient in creating conservative estimates for the FAP. However, as the observed SNR keeps on increasing the population of signals with positive log-Bayes factors decreases at the cost of the population with negative log-Bayes factors. Although at high SNR, the probability for a Gaussian process to have high log-Bayes factors is very small it is still non-zero. 

Figure~\ref{fig:SNRVar_MedianAlpha} and Figure ~\ref{fig:SNRVar_LogBayes} enable us to make statements on the distinctive character of our tool to differentiate a power-law GW signal and some unknown (terrestrial) correlated noise sources, simulated by the Gaussian process.

\subsection{Detection probability curve}

\begin{figure}
\centering
\includegraphics[width=0.48\textwidth]{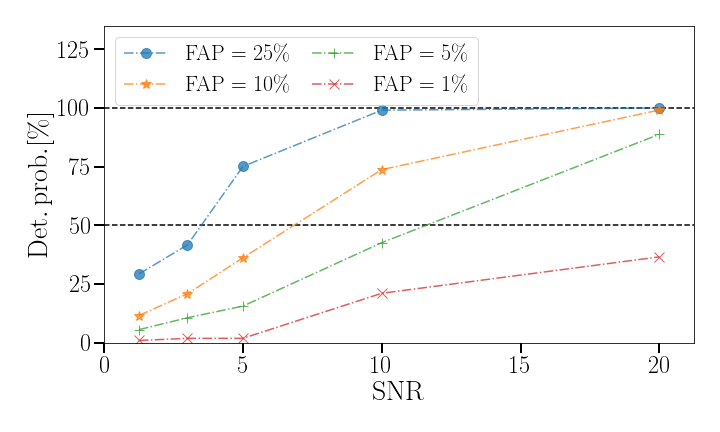}
\caption{Detection probability of a 2/3-power-law signal, for false alarm probabilities of 1\% (red), 5\%(green), 10\%(orange) and 25\% (blue). The black dashed lines indicate a detection probability of 50\% and 100\%. Signals were injected at SNR = 1.25, 3, 5, 10 and 20. For each signal type and SNR 5000 injections were performed.}
\label{fig:DetProbCurve}
\end{figure}

We construct a detection probability curve -- for the situation $\alpha = 2/3$ -- where we show the behaviour of the detection probability versus the SNR of the signal for several fixed false alarm rates. Injections at five different SNRs were performed: 1.25, 3, 5, 10 and 20. At each SNR, we performed 5000 injections for both the conservative Gaussian process signal as well as an $\alpha=2/3$ power-law signal. The result is shown in Fig.~\ref{fig:DetProbCurve}. 

The results indicate our Gaussian process is very conservative. Given a FAP of 1\%, the detection probability at SNR 20 is not even reaching 40\%. A detection probability of 50\% and 100\% is indicated by the black dashed curves. Given a FAP of 5\%, a detection probability of 50\% is reached above SNR=10. Only for a FAP of 25\%, this is reached for a SNR$\leq$5.\\
As shown in Table \ref{tab:FAP/FDP-SNR5}, in case of a power-law with SNR = 5 and a steeper slope, e.g. $\alpha = 3$, a detection probability curve of 41.40\% (67.08\%) is reached for a FAP of 1\% (5\%). This seems to indicate that even with our very conservative Gaussian process signal generation, a GWB signal with steeper power-law slope is more easily distinguishable from correlated terrestrial noise, here modeled by the Gaussian process.

\subsection{Application to a real life scenario - O2 outlier}
\label{sec:O2outlier}

\begin{figure}
\centering
\includegraphics[width=0.48\textwidth]{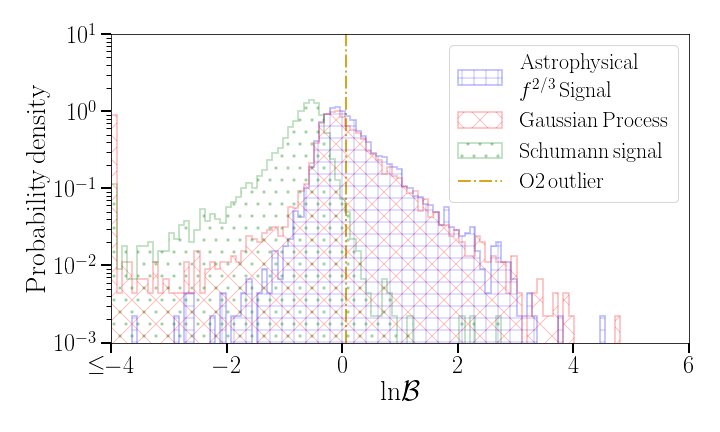}
\caption{Probability density of ln$\mathcal{B}$ for an astrophysical $\alpha=2/3$ power-law signal, random draws from a Gaussian process using our optimized (worst-case) SE-kernel and a Schumann signal. The lowest bin contains all ln$\mathcal{B} \leq -4$. For each signal type 5000 injections were performed with an injection strength of SNR=1.25. The choice of $\alpha=2/3$ and SNR=1.25 matches the parameters of observed outlier of the O2 results for an isotropic GWB\cite{O2Isotropic}.}
\label{fig:SNR1.25_Hist}
\end{figure}

When analyzing the results of their second observing run (O2), the LIGO and Virgo collaborations observed an excess of SNR 1.25 for a power-law model with $\alpha = 2/3$, as well as $\alpha = 3$ \citep{O2Isotropic}. At the time of the paper, it was stated the low SNR excess was very likely due to random fluctuations in the data. This is confirmed by the lack of detection by the subsequent O3 results \citep{O3Isotropic}. With the Geodesy tool described in this paper, we could, at the time of the O2 observation, have answered a complementary question: given that we have observed an excess with SNR=1.25 for a power-law model with $\alpha =2/3$, what is the probability the observed signal is due to a source of correlated noise instead of gravitational waves. Although the O3 results have confirmed the excess was just a random fluctuation, it is instructive to demonstrate how the Geodesy tool could be used in the future. In our demonstration, we will only investigate the excess given a power-law with $\alpha = 2/3$, although one can easily apply the tool to the $\alpha = 3$ case as well. \\
In what follows we will use the public available cross-correlation spectrum observed by LIGO during O2 \cite{publicO2} to compute the logBayes-factor linked to this observation. To construct the FAP and detection probability, we performed injections of the conservative Gaussian process signal, Schumann resonances, as well as a 2/3-power-law signal with an observed injection strength of SNR=1.25. All injections consist of 5000 samples and result in the distribution of log-Bayes factors shown in Fig.~\ref{fig:SNR1.25_Hist}.\\

The log-Bayes factor for the observed signal of the O2 run by LIGO and Virgo was computed and found to be 0.063. Given the distributions from the simulations shown in Fig.~\ref{fig:SNR1.25_Hist}, the observed signal is consistent with a FAP of $39.00\% \pm 0.02\%$ and a detection probability of $46.80\% \pm 0.02\%$. The high FAP does not give us enough confidence to prefer a gravitational-wave signal over a correlated noise source.\\
However if one compares a gravitational-wave signal with a correlated signal coming from Schumann resonances a FAP of $1.00\% \pm 0.02\%$ is found, effectively ruling out Schumann resonances as possible source with high confidence. This is consistent with projections showing there was no significant magnetic coupling in the analysis for an isotropic GWB using O2 data\cite{O2Isotropic}.

\section{Conclusion and Outlook}
\label{sec:DiscussionOutlook}
	
In this paper, we presented a tool that requires the observed signal to be consistent with the geometry of the observing detectors. We use Gaussian processes as a conservative proxy for the unknown space of all terrestrial correlated signals that might impact stochastic searches. This enables us to make quantitative statistical statements and false alarm probabilities concerning the origin of the observed signal. The framework was applied to a SNR=1.25 excess  for a 2/3-power law, observed by the LIGO and Virgo collaborations during their second observing run. Based on this analysis there was not enough evidence to prefer a GWB-signal over terrestrial correlated noise. However, Schumann resonances were effectively ruled out as possible source.\\ 
In this section we will discuss how this tool can be used in the future as well as the assumptions used in the current work and the possibilities for future improvements. \\

The primary use for this tool is when in the future (a hint of) a power-law isotropic GWB is observed. From analyzing the data three estimated parameters will be needed as input for the GW-Geodesy tool we are describing here: the power-law index $\alpha$, the observed SNR and the log-Bayes factor between \(\mathcal{H}_{\mathrm{Free}}\) and \(\mathcal{H}_{\gamma}\) for the observed signal. The observed $\alpha$ will be used to re-optimize the kernel parameters to get the worst case scenario for this specific power-law. The SNR will dictate the injection strength of our data. This will lead to a figure equivalent to Fig. \ref{fig:SNR3_Hist} in this work and log-Bayes factors linked to a certain FAP as in Tab. \ref{tab:FAP/FDP}. If the observed log-Bayes factor is larger or equal than the log-Bayes factor linked to the desired FAP$_{desired}$ (fixed beforehand) we can state the observed signal is preferred to come from gravitational waves instead of a source of correlated noise with a confidence of 1-FAP$_{observation}$. \\

Both in its current and previous form \citep{GeodesyOriginal}, the GW-Geodesy tool can only be used to validate an isotropic GWB. Currently it is being investigated how this tool can be extended to become applicable for an anisotropic GWB.\\
In this paper we only demonstrated the tool for the HL-baseline, since this is currently the most sensitive detector-pair for an isotropic GWB. It is however very easy to apply this technique to any preferred detector pair. At some point the ever increasing detector sensitivities and long observation times will make it possible a GWB will be observed by more than one detector pair. The current tool is able to make statements on all the baselines separately, but one can imagine extending the tool to get one overall figure of merit to make statements for the entire detector network.\\
It is important to note that the framework demonstrated in this paper is only tested when there is either a power-law GW signal present or a globally coherent noise source. The separation between a true GW signal and correlated noise becomes less straightforward if they are both present at the same time with similar strengths. In case of a known background a technique as proposed in \citep{BayesianGW-Mag} could be used to search for both sources at the same time.\\
One could also think of not only looking for power-law signals but more complex models. In the earlier implementation of the GW-Geodesy framework \citep{GeodesyOriginal} it was shown that the tool is quite robust against modeling a broken power-law with a single power-law. This could mean the framework is mainly sensitive to the zero crossings of the overlap reduction function.\\

\acknowledgements

The authors acknowledge access to computational resources provided by the LIGO Laboratory supported by National Science Foundation Grants PHY-0757058 and PHY-0823459.

This paper has been given LIGO DCC number P2100383 and Virgo TDS number VIR-1126A-21.

Kamiel Janssens is supported by FWO-Vlaanderen via grant number 11C5720N.
M.~W.~C acknowledges support from the National Science Foundation with grant number PHY-2010970.

\bibliographystyle{apsrev4-1}
\bibliography{references}

\begin{thebibliography}{41}%
\makeatletter
\providecommand \@ifxundefined [1]{%
 \@ifx{#1\undefined}
}%
\providecommand \@ifnum [1]{%
 \ifnum #1\expandafter \@firstoftwo
 \else \expandafter \@secondoftwo
 \fi
}%
\providecommand \@ifx [1]{%
 \ifx #1\expandafter \@firstoftwo
 \else \expandafter \@secondoftwo
 \fi
}%
\providecommand \natexlab [1]{#1}%
\providecommand \enquote  [1]{``#1''}%
\providecommand \bibnamefont  [1]{#1}%
\providecommand \bibfnamefont [1]{#1}%
\providecommand \citenamefont [1]{#1}%
\providecommand \href@noop [0]{\@secondoftwo}%
\providecommand \href [0]{\begingroup \@sanitize@url \@href}%
\providecommand \@href[1]{\@@startlink{#1}\@@href}%
\providecommand \@@href[1]{\endgroup#1\@@endlink}%
\providecommand \@sanitize@url [0]{\catcode `\\12\catcode `\$12\catcode
  `\&12\catcode `\#12\catcode `\^12\catcode `\_12\catcode `\%12\relax}%
\providecommand \@@startlink[1]{}%
\providecommand \@@endlink[0]{}%
\providecommand \url  [0]{\begingroup\@sanitize@url \@url }%
\providecommand \@url [1]{\endgroup\@href {#1}{\urlprefix }}%
\providecommand \urlprefix  [0]{URL }%
\providecommand \Eprint [0]{\href }%
\providecommand \doibase [0]{http://dx.doi.org/}%
\providecommand \selectlanguage [0]{\@gobble}%
\providecommand \bibinfo  [0]{\@secondoftwo}%
\providecommand \bibfield  [0]{\@secondoftwo}%
\providecommand \translation [1]{[#1]}%
\providecommand \BibitemOpen [0]{}%
\providecommand \bibitemStop [0]{}%
\providecommand \bibitemNoStop [0]{.\EOS\space}%
\providecommand \EOS [0]{\spacefactor3000\relax}%
\providecommand \BibitemShut  [1]{\csname bibitem#1\endcsname}%
\let\auto@bib@innerbib\@empty
\bibitem [{\citenamefont {Abbott}\ \emph {et~al.}(2016)\citenamefont {Abbott}
  \emph {et~al.}}]{FirstGWObs}%
  \BibitemOpen
  \bibfield  {author} {\bibinfo {author} {\bibfnamefont {B.~P.}\ \bibnamefont
  {Abbott}} \emph {et~al.} (\bibinfo {collaboration} {LIGO Scientific
  Collaboration and Virgo Collaboration}),\ }\href {\doibase
  10.1103/PhysRevLett.116.061102} {\bibfield  {journal} {\bibinfo  {journal}
  {Phys. Rev. Lett.}\ }\textbf {\bibinfo {volume} {116}},\ \bibinfo {pages}
  {061102} (\bibinfo {year} {2016})}\BibitemShut {NoStop}%
\bibitem [{\citenamefont {Aasi}\ \emph {et~al.}(2015)\citenamefont {Aasi} \emph
  {et~al.}}]{2015}%
  \BibitemOpen
  \bibfield  {author} {\bibinfo {author} {\bibfnamefont {J.}~\bibnamefont
  {Aasi}} \emph {et~al.},\ }\href {\doibase 10.1088/0264-9381/32/7/074001}
  {\bibfield  {journal} {\bibinfo  {journal} {Classical and Quantum Gravity}\
  }\textbf {\bibinfo {volume} {32}},\ \bibinfo {pages} {074001} (\bibinfo
  {year} {2015})}\BibitemShut {NoStop}%
\bibitem [{\citenamefont {Acernese}\ \emph {et~al.}(2015)\citenamefont
  {Acernese} \emph {et~al.}}]{VIRGO:2014yos}%
  \BibitemOpen
  \bibfield  {author} {\bibinfo {author} {\bibfnamefont {F.}~\bibnamefont
  {Acernese}} \emph {et~al.} (\bibinfo {collaboration} {VIRGO}),\ }\href
  {\doibase 10.1088/0264-9381/32/2/024001} {\bibfield  {journal} {\bibinfo
  {journal} {Class. Quant. Grav.}\ }\textbf {\bibinfo {volume} {32}},\ \bibinfo
  {pages} {024001} (\bibinfo {year} {2015})},\ \Eprint
  {http://arxiv.org/abs/1408.3978} {arXiv:1408.3978 [gr-qc]} \BibitemShut
  {NoStop}%
\bibitem [{\citenamefont {Aso}\ \emph {et~al.}(2013)\citenamefont {Aso},
  \citenamefont {Michimura}, \citenamefont {Somiya}, \citenamefont {Ando},
  \citenamefont {Miyakawa}, \citenamefont {Sekiguchi}, \citenamefont
  {Tatsumi},\ and\ \citenamefont {Yamamoto}}]{PhysRevD.88.043007}%
  \BibitemOpen
  \bibfield  {author} {\bibinfo {author} {\bibfnamefont {Y.}~\bibnamefont
  {Aso}}, \bibinfo {author} {\bibfnamefont {Y.}~\bibnamefont {Michimura}},
  \bibinfo {author} {\bibfnamefont {K.}~\bibnamefont {Somiya}}, \bibinfo
  {author} {\bibfnamefont {M.}~\bibnamefont {Ando}}, \bibinfo {author}
  {\bibfnamefont {O.}~\bibnamefont {Miyakawa}}, \bibinfo {author}
  {\bibfnamefont {T.}~\bibnamefont {Sekiguchi}}, \bibinfo {author}
  {\bibfnamefont {D.}~\bibnamefont {Tatsumi}}, \ and\ \bibinfo {author}
  {\bibfnamefont {H.}~\bibnamefont {Yamamoto}} (\bibinfo {collaboration} {The
  KAGRA Collaboration}),\ }\href {\doibase 10.1103/PhysRevD.88.043007}
  {\bibfield  {journal} {\bibinfo  {journal} {Phys. Rev. D}\ }\textbf {\bibinfo
  {volume} {88}},\ \bibinfo {pages} {043007} (\bibinfo {year}
  {2013})}\BibitemShut {NoStop}%
\bibitem [{\citenamefont {Abbott}\ \emph
  {et~al.}(2019{\natexlab{a}})\citenamefont {Abbott} \emph {et~al.}}]{GWTC1}%
  \BibitemOpen
  \bibfield  {author} {\bibinfo {author} {\bibfnamefont {B.~P.}\ \bibnamefont
  {Abbott}} \emph {et~al.} (\bibinfo {collaboration} {LIGO Scientific
  Collaboration and Virgo Collaboration}),\ }\href {\doibase
  10.1103/PhysRevX.9.031040} {\bibfield  {journal} {\bibinfo  {journal} {Phys.
  Rev. X}\ }\textbf {\bibinfo {volume} {9}},\ \bibinfo {pages} {031040}
  (\bibinfo {year} {2019}{\natexlab{a}})}\BibitemShut {NoStop}%
\bibitem [{\citenamefont {{Abbott}}\ \emph {et~al.}(2020)\citenamefont
  {{Abbott}} \emph {et~al.}}]{GWTC2}%
  \BibitemOpen
  \bibfield  {author} {\bibinfo {author} {\bibfnamefont {R.}~\bibnamefont
  {{Abbott}}} \emph {et~al.},\ }\href@noop {} {\bibfield  {journal} {\bibinfo
  {journal} {arXiv e-prints}\ ,\ \bibinfo {eid} {arXiv:2010.14527}} (\bibinfo
  {year} {2020})},\ \Eprint {http://arxiv.org/abs/2010.14527} {arXiv:2010.14527
  [gr-qc]} \BibitemShut {NoStop}%
\bibitem [{\citenamefont {Abbott}\ \emph
  {et~al.}(2021{\natexlab{a}})\citenamefont {Abbott}, \citenamefont {Abbott},\
  and\ \citenamefont {et~al}}]{NSBHDiscovery}%
  \BibitemOpen
  \bibfield  {author} {\bibinfo {author} {\bibfnamefont {R.}~\bibnamefont
  {Abbott}}, \bibinfo {author} {\bibfnamefont {T.~D.}\ \bibnamefont {Abbott}},
  \ and\ \bibinfo {author} {\bibnamefont {et~al}},\ }\href {\doibase
  10.3847/2041-8213/ac082e} {\bibfield  {journal} {\bibinfo  {journal} {The
  Astrophysical Journal Letters}\ }\textbf {\bibinfo {volume} {915}},\ \bibinfo
  {pages} {L5} (\bibinfo {year} {2021}{\natexlab{a}})}\BibitemShut {NoStop}%
\bibitem [{\citenamefont {Collaboration}\ \emph {et~al.}(2021)\citenamefont
  {Collaboration}, \citenamefont {the Virgo~Collaboration}, \citenamefont {the
  KAGRA~Collaboration}, \citenamefont {Abbott}, \citenamefont {Abbott},
  \citenamefont {Acernese}, \citenamefont {Ackley},\ and\ \citenamefont
  {et~al}}]{GWTC3}%
  \BibitemOpen
  \bibfield  {author} {\bibinfo {author} {\bibfnamefont {T.~L.~S.}\
  \bibnamefont {Collaboration}}, \bibinfo {author} {\bibnamefont {the
  Virgo~Collaboration}}, \bibinfo {author} {\bibnamefont {the
  KAGRA~Collaboration}}, \bibinfo {author} {\bibfnamefont {R.}~\bibnamefont
  {Abbott}}, \bibinfo {author} {\bibfnamefont {T.~D.}\ \bibnamefont {Abbott}},
  \bibinfo {author} {\bibfnamefont {F.}~\bibnamefont {Acernese}}, \bibinfo
  {author} {\bibfnamefont {K.}~\bibnamefont {Ackley}}, \ and\ \bibinfo {author}
  {\bibnamefont {et~al}},\ }\href@noop {} {\enquote {\bibinfo {title} {Gwtc-3:
  Compact binary coalescences observed by ligo and virgo during the second part
  of the third observing run},}\ } (\bibinfo {year} {2021}),\ \Eprint
  {http://arxiv.org/abs/2111.03606} {arXiv:2111.03606 [gr-qc]} \BibitemShut
  {NoStop}%
\bibitem [{\citenamefont {Abbott}\ \emph
  {et~al.}(2021{\natexlab{b}})\citenamefont {Abbott} \emph {et~al.}}]{GWTC2.1}%
  \BibitemOpen
  \bibfield  {author} {\bibinfo {author} {\bibfnamefont {R.}~\bibnamefont
  {Abbott}} \emph {et~al.} (\bibinfo {collaboration} {The LIGO Scientific
  Collaboration and the Virgo Collaboration}),\ }\href@noop {} {\enquote
  {\bibinfo {title} {Gwtc-2.1: Deep extended catalog of compact binary
  coalescences observed by ligo and virgo during the first half of the third
  observing run},}\ } (\bibinfo {year} {2021}{\natexlab{b}}),\ \Eprint
  {http://arxiv.org/abs/2108.01045} {arXiv:2108.01045 [gr-qc]} \BibitemShut
  {NoStop}%
\bibitem [{\citenamefont {Regimbau}\ and\ \citenamefont
  {Mandic}(2008)}]{Regimbau_2008}%
  \BibitemOpen
  \bibfield  {author} {\bibinfo {author} {\bibfnamefont {T.}~\bibnamefont
  {Regimbau}}\ and\ \bibinfo {author} {\bibfnamefont {V.}~\bibnamefont
  {Mandic}},\ }\href {\doibase 10.1088/0264-9381/25/18/184018} {\bibfield
  {journal} {\bibinfo  {journal} {Classical and Quantum Gravity}\ }\textbf
  {\bibinfo {volume} {25}},\ \bibinfo {pages} {184018} (\bibinfo {year}
  {2008})}\BibitemShut {NoStop}%
\bibitem [{\citenamefont {Rosado}(2011)}]{CBC-GWB-1}%
  \BibitemOpen
  \bibfield  {author} {\bibinfo {author} {\bibfnamefont {P.~A.}\ \bibnamefont
  {Rosado}},\ }\href {\doibase 10.1103/PhysRevD.84.084004} {\bibfield
  {journal} {\bibinfo  {journal} {Phys. Rev. D}\ }\textbf {\bibinfo {volume}
  {84}},\ \bibinfo {pages} {084004} (\bibinfo {year} {2011})}\BibitemShut
  {NoStop}%
\bibitem [{\citenamefont {Zhu}\ \emph {et~al.}(2011)\citenamefont {Zhu},
  \citenamefont {Howell}, \citenamefont {Regimbau}, \citenamefont {Blair},\
  and\ \citenamefont {Zhu}}]{CBC-GWB-2}%
  \BibitemOpen
  \bibfield  {author} {\bibinfo {author} {\bibfnamefont {X.-J.}\ \bibnamefont
  {Zhu}}, \bibinfo {author} {\bibfnamefont {E.}~\bibnamefont {Howell}},
  \bibinfo {author} {\bibfnamefont {T.}~\bibnamefont {Regimbau}}, \bibinfo
  {author} {\bibfnamefont {D.}~\bibnamefont {Blair}}, \ and\ \bibinfo {author}
  {\bibfnamefont {Z.-H.}\ \bibnamefont {Zhu}},\ }\href {\doibase
  10.1088/0004-637x/739/2/86} {\bibfield  {journal} {\bibinfo  {journal} {The
  Astrophysical Journal}\ }\textbf {\bibinfo {volume} {739}},\ \bibinfo {pages}
  {86} (\bibinfo {year} {2011})}\BibitemShut {NoStop}%
\bibitem [{\citenamefont {Marassi}\ \emph {et~al.}(2011)\citenamefont
  {Marassi}, \citenamefont {Schneider}, \citenamefont {Corvino}, \citenamefont
  {Ferrari},\ and\ \citenamefont {Zwart}}]{CBC-GWB-3}%
  \BibitemOpen
  \bibfield  {author} {\bibinfo {author} {\bibfnamefont {S.}~\bibnamefont
  {Marassi}}, \bibinfo {author} {\bibfnamefont {R.}~\bibnamefont {Schneider}},
  \bibinfo {author} {\bibfnamefont {G.}~\bibnamefont {Corvino}}, \bibinfo
  {author} {\bibfnamefont {V.}~\bibnamefont {Ferrari}}, \ and\ \bibinfo
  {author} {\bibfnamefont {S.~P.}\ \bibnamefont {Zwart}},\ }\href {\doibase
  10.1103/PhysRevD.84.124037} {\bibfield  {journal} {\bibinfo  {journal} {Phys.
  Rev. D}\ }\textbf {\bibinfo {volume} {84}},\ \bibinfo {pages} {124037}
  (\bibinfo {year} {2011})}\BibitemShut {NoStop}%
\bibitem [{\citenamefont {Wu}\ \emph {et~al.}(2012)\citenamefont {Wu},
  \citenamefont {Mandic},\ and\ \citenamefont {Regimbau}}]{CBC-GWB-4}%
  \BibitemOpen
  \bibfield  {author} {\bibinfo {author} {\bibfnamefont {C.}~\bibnamefont
  {Wu}}, \bibinfo {author} {\bibfnamefont {V.}~\bibnamefont {Mandic}}, \ and\
  \bibinfo {author} {\bibfnamefont {T.}~\bibnamefont {Regimbau}},\ }\href
  {\doibase 10.1103/PhysRevD.85.104024} {\bibfield  {journal} {\bibinfo
  {journal} {Phys. Rev. D}\ }\textbf {\bibinfo {volume} {85}},\ \bibinfo
  {pages} {104024} (\bibinfo {year} {2012})}\BibitemShut {NoStop}%
\bibitem [{\citenamefont {Zhu}\ \emph {et~al.}(2013)\citenamefont {Zhu},
  \citenamefont {Howell}, \citenamefont {Blair},\ and\ \citenamefont
  {Zhu}}]{CBC-GWB-5}%
  \BibitemOpen
  \bibfield  {author} {\bibinfo {author} {\bibfnamefont {X.-J.}\ \bibnamefont
  {Zhu}}, \bibinfo {author} {\bibfnamefont {E.~J.}\ \bibnamefont {Howell}},
  \bibinfo {author} {\bibfnamefont {D.~G.}\ \bibnamefont {Blair}}, \ and\
  \bibinfo {author} {\bibfnamefont {Z.-H.}\ \bibnamefont {Zhu}},\ }\href
  {\doibase 10.1093/mnras/stt207} {\bibfield  {journal} {\bibinfo  {journal}
  {Monthly Notices of the Royal Astronomical Society}\ }\textbf {\bibinfo
  {volume} {431}},\ \bibinfo {pages} {882} (\bibinfo {year} {2013})},\ \Eprint
  {http://arxiv.org/abs/https://academic.oup.com/mnras/article-pdf/431/1/882/18243620/stt207.pdf}
  {https://academic.oup.com/mnras/article-pdf/431/1/882/18243620/stt207.pdf}
  \BibitemShut {NoStop}%
\bibitem [{\citenamefont {Abbott}\ \emph
  {et~al.}(2021{\natexlab{c}})\citenamefont {Abbott} \emph
  {et~al.}}]{O3Isotropic}%
  \BibitemOpen
  \bibfield  {author} {\bibinfo {author} {\bibfnamefont {R.}~\bibnamefont
  {Abbott}} \emph {et~al.} (\bibinfo {collaboration} {LIGO Scientific
  Collaboration, Virgo Collaboration, and KAGRA Collaboration}),\ }\href
  {\doibase 10.1103/PhysRevD.104.022004} {\bibfield  {journal} {\bibinfo
  {journal} {Phys. Rev. D}\ }\textbf {\bibinfo {volume} {104}},\ \bibinfo
  {pages} {022004} (\bibinfo {year} {2021}{\natexlab{c}})}\BibitemShut
  {NoStop}%
\bibitem [{\citenamefont {Abbott}\ \emph
  {et~al.}(2021{\natexlab{d}})\citenamefont {Abbott} \emph
  {et~al.}}]{O3Directional}%
  \BibitemOpen
  \bibfield  {author} {\bibinfo {author} {\bibfnamefont {R.}~\bibnamefont
  {Abbott}} \emph {et~al.} (\bibinfo {collaboration} {LIGO Scientific
  Collaboration, Virgo Collaboration, and KAGRA Collaboration}),\ }\href
  {\doibase 10.1103/PhysRevD.104.022005} {\bibfield  {journal} {\bibinfo
  {journal} {Phys. Rev. D}\ }\textbf {\bibinfo {volume} {104}},\ \bibinfo
  {pages} {022005} (\bibinfo {year} {2021}{\natexlab{d}})}\BibitemShut
  {NoStop}%
\bibitem [{\citenamefont {Abbott}\ \emph
  {et~al.}(2020{\natexlab{a}})\citenamefont {Abbott}, \citenamefont {Abbott},
  \citenamefont {Abbott}, \citenamefont {Abraham}, \citenamefont {Acernese},
  \citenamefont {Ackley}, \citenamefont {Adams}, \citenamefont {Adya},
  \citenamefont {Affeldt}, \citenamefont {Agathos},\ and\ \citenamefont
  {et~al.}}]{ObsProspects}%
  \BibitemOpen
  \bibfield  {author} {\bibinfo {author} {\bibfnamefont {B.~P.}\ \bibnamefont
  {Abbott}}, \bibinfo {author} {\bibfnamefont {R.}~\bibnamefont {Abbott}},
  \bibinfo {author} {\bibfnamefont {T.~D.}\ \bibnamefont {Abbott}}, \bibinfo
  {author} {\bibfnamefont {S.}~\bibnamefont {Abraham}}, \bibinfo {author}
  {\bibfnamefont {F.}~\bibnamefont {Acernese}}, \bibinfo {author}
  {\bibfnamefont {K.}~\bibnamefont {Ackley}}, \bibinfo {author} {\bibfnamefont
  {C.}~\bibnamefont {Adams}}, \bibinfo {author} {\bibfnamefont {V.~B.}\
  \bibnamefont {Adya}}, \bibinfo {author} {\bibfnamefont {C.}~\bibnamefont
  {Affeldt}}, \bibinfo {author} {\bibfnamefont {M.}~\bibnamefont {Agathos}}, \
  and\ \bibinfo {author} {\bibnamefont {et~al.}},\ }\href {\doibase
  10.1007/s41114-020-00026-9} {\bibfield  {journal} {\bibinfo  {journal}
  {Living Reviews in Relativity}\ }\textbf {\bibinfo {volume} {23}} (\bibinfo
  {year} {2020}{\natexlab{a}}),\ 10.1007/s41114-020-00026-9}\BibitemShut
  {NoStop}%
\bibitem [{\citenamefont {Christensen}(2018)}]{SGWBRevChristensen}%
  \BibitemOpen
  \bibfield  {author} {\bibinfo {author} {\bibfnamefont {N.}~\bibnamefont
  {Christensen}},\ }\href {\doibase 10.1088/1361-6633/aae6b5} {\bibfield
  {journal} {\bibinfo  {journal} {Reports on Progress in Physics}\ }\textbf
  {\bibinfo {volume} {82}},\ \bibinfo {pages} {016903} (\bibinfo {year}
  {2018})}\BibitemShut {NoStop}%
\bibitem [{\citenamefont {Romano}\ and\ \citenamefont
  {Cornish}(2017)}]{SGWBRevRomano}%
  \BibitemOpen
  \bibfield  {author} {\bibinfo {author} {\bibfnamefont {J.~D.}\ \bibnamefont
  {Romano}}\ and\ \bibinfo {author} {\bibfnamefont {N.~J.}\ \bibnamefont
  {Cornish}},\ }\href {\doibase 10.1007/s41114-017-0004-1} {\bibfield
  {journal} {\bibinfo  {journal} {Living Rev. Relativ.}\ }\textbf {\bibinfo
  {volume} {20}},\ \bibinfo {pages} {2} (\bibinfo {year} {2017})},\ \Eprint
  {http://arxiv.org/abs/1608.06889} {arXiv:1608.06889} \BibitemShut {NoStop}%
\bibitem [{\citenamefont {{Schumann}}(1952{\natexlab{a}})}]{Schumann1}%
  \BibitemOpen
  \bibfield  {author} {\bibinfo {author} {\bibfnamefont {W.~O.}\ \bibnamefont
  {{Schumann}}},\ }\href {\doibase 10.1515/zna-1952-0202} {\bibfield  {journal}
  {\bibinfo  {journal} {Zeitschrift Naturforschung Teil A}\ }\textbf {\bibinfo
  {volume} {7}},\ \bibinfo {pages} {149} (\bibinfo {year}
  {1952}{\natexlab{a}})}\BibitemShut {NoStop}%
\bibitem [{\citenamefont {{Schumann}}(1952{\natexlab{b}})}]{Schumann2}%
  \BibitemOpen
  \bibfield  {author} {\bibinfo {author} {\bibfnamefont {W.~O.}\ \bibnamefont
  {{Schumann}}},\ }\href {\doibase 10.1515/zna-1952-3-404} {\bibfield
  {journal} {\bibinfo  {journal} {Zeitschrift Naturforschung Teil A}\ }\textbf
  {\bibinfo {volume} {7}},\ \bibinfo {pages} {250} (\bibinfo {year}
  {1952}{\natexlab{b}})}\BibitemShut {NoStop}%
\bibitem [{\citenamefont {{Covas}}\ \emph {et~al.}(2018)\citenamefont
  {{Covas}}, \citenamefont {{Effler}}, \citenamefont {{Goetz}}, \citenamefont
  {{Meyers}}, \citenamefont {{Neunzert}}, \citenamefont {{Oliver}},
  \citenamefont {{Pearlstone}}, \citenamefont {{Roma}}, \citenamefont
  {{Schofield}}, \citenamefont {{Adya}},\ and\ \citenamefont
  {et~al.}}]{Covas2018}%
  \BibitemOpen
  \bibfield  {author} {\bibinfo {author} {\bibfnamefont {P.~B.}\ \bibnamefont
  {{Covas}}}, \bibinfo {author} {\bibfnamefont {A.}~\bibnamefont {{Effler}}},
  \bibinfo {author} {\bibfnamefont {E.}~\bibnamefont {{Goetz}}}, \bibinfo
  {author} {\bibfnamefont {P.~M.}\ \bibnamefont {{Meyers}}}, \bibinfo {author}
  {\bibfnamefont {A.}~\bibnamefont {{Neunzert}}}, \bibinfo {author}
  {\bibfnamefont {M.}~\bibnamefont {{Oliver}}}, \bibinfo {author}
  {\bibfnamefont {B.~L.}\ \bibnamefont {{Pearlstone}}}, \bibinfo {author}
  {\bibfnamefont {V.~J.}\ \bibnamefont {{Roma}}}, \bibinfo {author}
  {\bibfnamefont {R.~M.~S.}\ \bibnamefont {{Schofield}}}, \bibinfo {author}
  {\bibfnamefont {V.~B.}\ \bibnamefont {{Adya}}}, \ and\ \bibinfo {author}
  {\bibnamefont {et~al.}},\ }\href {\doibase 10.1103/PhysRevD.97.082002}
  {\bibfield  {journal} {\bibinfo  {journal} {\prd}\ }\textbf {\bibinfo
  {volume} {97}},\ \bibinfo {eid} {082002} (\bibinfo {year} {2018})},\ \Eprint
  {http://arxiv.org/abs/1801.07204} {arXiv:1801.07204 [astro-ph.IM]}
  \BibitemShut {NoStop}%
\bibitem [{\citenamefont {Thrane}\ \emph {et~al.}(2013)\citenamefont {Thrane},
  \citenamefont {Christensen},\ and\ \citenamefont {Schofield}}]{Thrane2013}%
  \BibitemOpen
  \bibfield  {author} {\bibinfo {author} {\bibfnamefont {E.}~\bibnamefont
  {Thrane}}, \bibinfo {author} {\bibfnamefont {N.}~\bibnamefont {Christensen}},
  \ and\ \bibinfo {author} {\bibfnamefont {R.~M.~S.}\ \bibnamefont
  {Schofield}},\ }\href {\doibase 10.1103/PhysRevD.87.123009} {\bibfield
  {journal} {\bibinfo  {journal} {Phys. Rev. D}\ }\textbf {\bibinfo {volume}
  {87}},\ \bibinfo {pages} {123009} (\bibinfo {year} {2013})},\ \Eprint
  {http://arxiv.org/abs/1303.2613} {arXiv:1303.2613} \BibitemShut {NoStop}%
\bibitem [{\citenamefont {Thrane}\ \emph {et~al.}(2014)\citenamefont {Thrane},
  \citenamefont {Christensen}, \citenamefont {Schofield},\ and\ \citenamefont
  {Effler}}]{Thrane2014}%
  \BibitemOpen
  \bibfield  {author} {\bibinfo {author} {\bibfnamefont {E.}~\bibnamefont
  {Thrane}}, \bibinfo {author} {\bibfnamefont {N.}~\bibnamefont {Christensen}},
  \bibinfo {author} {\bibfnamefont {R.~M.~S.}\ \bibnamefont {Schofield}}, \
  and\ \bibinfo {author} {\bibfnamefont {A.}~\bibnamefont {Effler}},\ }\href
  {\doibase 10.1103/PhysRevD.90.023013} {\bibfield  {journal} {\bibinfo
  {journal} {Phys. Rev. D}\ }\textbf {\bibinfo {volume} {90}},\ \bibinfo
  {pages} {023013} (\bibinfo {year} {2014})},\ \Eprint
  {http://arxiv.org/abs/1406.2367} {arXiv:1406.2367} \BibitemShut {NoStop}%
\bibitem [{\citenamefont {Coughlin}\ \emph {et~al.}(2016)\citenamefont
  {Coughlin} \emph {et~al.}}]{Coughlin2016}%
  \BibitemOpen
  \bibfield  {author} {\bibinfo {author} {\bibfnamefont {M.~W.}\ \bibnamefont
  {Coughlin}} \emph {et~al.},\ }\href {\doibase 10.1088/0264-9381/33/22/224003}
  {\bibfield  {journal} {\bibinfo  {journal} {Class. Quant. Grav.}\ }\textbf
  {\bibinfo {volume} {33}},\ \bibinfo {pages} {224003} (\bibinfo {year}
  {2016})},\ \Eprint {http://arxiv.org/abs/1606.01011} {arXiv:1606.01011
  [gr-qc]} \BibitemShut {NoStop}%
\bibitem [{\citenamefont {Coughlin}\ \emph {et~al.}(2018)\citenamefont
  {Coughlin}, \citenamefont {Cirone}, \citenamefont {Meyers}, \citenamefont
  {Atsuta}, \citenamefont {Boschi}, \citenamefont {Chincarini}, \citenamefont
  {Christensen}, \citenamefont {De~Rosa}, \citenamefont {Effler}, \citenamefont
  {Fiori}, \citenamefont {Go\l{}kowski}, \citenamefont {Guidry}, \citenamefont
  {Harms}, \citenamefont {Hayama}, \citenamefont {Kataoka}, \citenamefont
  {Kubisz}, \citenamefont {Kulak}, \citenamefont {Laxen}, \citenamefont
  {Matas}, \citenamefont {Mlynarczyk}, \citenamefont {Ogawa}, \citenamefont
  {Paoletti}, \citenamefont {Salvador}, \citenamefont {Schofield},
  \citenamefont {Somiya},\ and\ \citenamefont {Thrane}}]{Coughlin2018}%
  \BibitemOpen
  \bibfield  {author} {\bibinfo {author} {\bibfnamefont {M.~W.}\ \bibnamefont
  {Coughlin}}, \bibinfo {author} {\bibfnamefont {A.}~\bibnamefont {Cirone}},
  \bibinfo {author} {\bibfnamefont {P.}~\bibnamefont {Meyers}}, \bibinfo
  {author} {\bibfnamefont {S.}~\bibnamefont {Atsuta}}, \bibinfo {author}
  {\bibfnamefont {V.}~\bibnamefont {Boschi}}, \bibinfo {author} {\bibfnamefont
  {A.}~\bibnamefont {Chincarini}}, \bibinfo {author} {\bibfnamefont {N.~L.}\
  \bibnamefont {Christensen}}, \bibinfo {author} {\bibfnamefont
  {R.}~\bibnamefont {De~Rosa}}, \bibinfo {author} {\bibfnamefont
  {A.}~\bibnamefont {Effler}}, \bibinfo {author} {\bibfnamefont
  {I.}~\bibnamefont {Fiori}}, \bibinfo {author} {\bibfnamefont
  {M.}~\bibnamefont {Go\l{}kowski}}, \bibinfo {author} {\bibfnamefont
  {M.}~\bibnamefont {Guidry}}, \bibinfo {author} {\bibfnamefont
  {J.}~\bibnamefont {Harms}}, \bibinfo {author} {\bibfnamefont
  {K.}~\bibnamefont {Hayama}}, \bibinfo {author} {\bibfnamefont
  {Y.}~\bibnamefont {Kataoka}}, \bibinfo {author} {\bibfnamefont
  {J.}~\bibnamefont {Kubisz}}, \bibinfo {author} {\bibfnamefont
  {A.}~\bibnamefont {Kulak}}, \bibinfo {author} {\bibfnamefont
  {M.}~\bibnamefont {Laxen}}, \bibinfo {author} {\bibfnamefont
  {A.}~\bibnamefont {Matas}}, \bibinfo {author} {\bibfnamefont
  {J.}~\bibnamefont {Mlynarczyk}}, \bibinfo {author} {\bibfnamefont
  {T.}~\bibnamefont {Ogawa}}, \bibinfo {author} {\bibfnamefont
  {F.}~\bibnamefont {Paoletti}}, \bibinfo {author} {\bibfnamefont
  {J.}~\bibnamefont {Salvador}}, \bibinfo {author} {\bibfnamefont
  {R.}~\bibnamefont {Schofield}}, \bibinfo {author} {\bibfnamefont
  {K.}~\bibnamefont {Somiya}}, \ and\ \bibinfo {author} {\bibfnamefont
  {E.}~\bibnamefont {Thrane}},\ }\href {\doibase 10.1103/PhysRevD.97.102007}
  {\bibfield  {journal} {\bibinfo  {journal} {Phys. Rev. D}\ }\textbf {\bibinfo
  {volume} {97}},\ \bibinfo {pages} {102007} (\bibinfo {year}
  {2018})}\BibitemShut {NoStop}%
\bibitem [{\citenamefont {Meyers}\ \emph {et~al.}(2020)\citenamefont {Meyers},
  \citenamefont {Martinovic}, \citenamefont {Christensen},\ and\ \citenamefont
  {Sakellariadou}}]{BayesianGW-Mag}%
  \BibitemOpen
  \bibfield  {author} {\bibinfo {author} {\bibfnamefont {P.~M.}\ \bibnamefont
  {Meyers}}, \bibinfo {author} {\bibfnamefont {K.}~\bibnamefont {Martinovic}},
  \bibinfo {author} {\bibfnamefont {N.}~\bibnamefont {Christensen}}, \ and\
  \bibinfo {author} {\bibfnamefont {M.}~\bibnamefont {Sakellariadou}},\ }\href
  {\doibase 10.1103/PhysRevD.102.102005} {\bibfield  {journal} {\bibinfo
  {journal} {Physical Review D}\ }\textbf {\bibinfo {volume} {102}},\ \bibinfo
  {pages} {102005} (\bibinfo {year} {2020})}\BibitemShut {NoStop}%
\bibitem [{\citenamefont {Callister}\ \emph {et~al.}(2018)\citenamefont
  {Callister}, \citenamefont {Coughlin},\ and\ \citenamefont
  {Kanner}}]{GeodesyOriginal}%
  \BibitemOpen
  \bibfield  {author} {\bibinfo {author} {\bibfnamefont {T.~A.}\ \bibnamefont
  {Callister}}, \bibinfo {author} {\bibfnamefont {M.~W.}\ \bibnamefont
  {Coughlin}}, \ and\ \bibinfo {author} {\bibfnamefont {J.~B.}\ \bibnamefont
  {Kanner}},\ }\href {\doibase 10.3847/2041-8213/aaf3a5} {\bibfield  {journal}
  {\bibinfo  {journal} {The Astrophysical Journal}\ }\textbf {\bibinfo {volume}
  {869}},\ \bibinfo {pages} {L28} (\bibinfo {year} {2018})}\BibitemShut
  {NoStop}%
\bibitem [{\citenamefont {Allen}\ and\ \citenamefont
  {Romano}(1999)}]{SGWBRevAllen}%
  \BibitemOpen
  \bibfield  {author} {\bibinfo {author} {\bibfnamefont {B.}~\bibnamefont
  {Allen}}\ and\ \bibinfo {author} {\bibfnamefont {J.~D.}\ \bibnamefont
  {Romano}},\ }\href {\doibase 10.1103/PhysRevD.59.102001} {\bibfield
  {journal} {\bibinfo  {journal} {Phys. Rev. D}\ }\textbf {\bibinfo {volume}
  {59}},\ \bibinfo {pages} {102001} (\bibinfo {year} {1999})}\BibitemShut
  {NoStop}%
\bibitem [{\citenamefont {Christensen}(1992)}]{SGWBRevChristensen1992}%
  \BibitemOpen
  \bibfield  {author} {\bibinfo {author} {\bibfnamefont {N.}~\bibnamefont
  {Christensen}},\ }\href {\doibase 10.1103/PhysRevD.46.5250} {\bibfield
  {journal} {\bibinfo  {journal} {Phys. Rev. D}\ }\textbf {\bibinfo {volume}
  {46}},\ \bibinfo {pages} {5250} (\bibinfo {year} {1992})}\BibitemShut
  {NoStop}%
\bibitem [{\citenamefont {Regimbau}(2011)}]{SGWBTania}%
  \BibitemOpen
  \bibfield  {author} {\bibinfo {author} {\bibfnamefont {T.}~\bibnamefont
  {Regimbau}},\ }\href {\doibase 10.1088/1674-4527/11/4/001} {\ \textbf
  {\bibinfo {volume} {11}},\ \bibinfo {pages} {369} (\bibinfo {year}
  {2011})}\BibitemShut {NoStop}%
\bibitem [{\citenamefont {Rasmussen}\ and\ \citenamefont
  {Williams}(2006)}]{GPForMachineLearning}%
  \BibitemOpen
  \bibfield  {author} {\bibinfo {author} {\bibfnamefont {C.~E.}\ \bibnamefont
  {Rasmussen}}\ and\ \bibinfo {author} {\bibfnamefont {C.~K.~I.}\ \bibnamefont
  {Williams}},\ }\href {http://www.gaussianprocess.org/gpml/chapters/RW.pdf}
  {\emph {\bibinfo {title} {Gaussian Processes for Machine Learning}}}\
  (\bibinfo  {publisher} {The MIT Press},\ \bibinfo {year} {2006})\BibitemShut
  {NoStop}%
\bibitem [{\citenamefont {Abbott}\ \emph
  {et~al.}(2020{\natexlab{b}})\citenamefont {Abbott}, \citenamefont {Abbott},
  \citenamefont {Abbott}, \citenamefont {Abraham}, \citenamefont {Acernese},
  \citenamefont {Ackley}, \citenamefont {Adams}, \citenamefont {Adya},
  \citenamefont {Affeldt}, \citenamefont {Agathos},\ and\ \citenamefont
  {et~al.}}]{LIGOSensitivity}%
  \BibitemOpen
  \bibfield  {author} {\bibinfo {author} {\bibfnamefont {B.~P.}\ \bibnamefont
  {Abbott}}, \bibinfo {author} {\bibfnamefont {R.}~\bibnamefont {Abbott}},
  \bibinfo {author} {\bibfnamefont {T.~D.}\ \bibnamefont {Abbott}}, \bibinfo
  {author} {\bibfnamefont {S.}~\bibnamefont {Abraham}}, \bibinfo {author}
  {\bibfnamefont {F.}~\bibnamefont {Acernese}}, \bibinfo {author}
  {\bibfnamefont {K.}~\bibnamefont {Ackley}}, \bibinfo {author} {\bibfnamefont
  {C.}~\bibnamefont {Adams}}, \bibinfo {author} {\bibfnamefont {V.~B.}\
  \bibnamefont {Adya}}, \bibinfo {author} {\bibfnamefont {C.}~\bibnamefont
  {Affeldt}}, \bibinfo {author} {\bibfnamefont {M.}~\bibnamefont {Agathos}}, \
  and\ \bibinfo {author} {\bibnamefont {et~al.}},\ }\href {\doibase
  10.1007/s41114-020-00026-9} {\bibfield  {journal} {\bibinfo  {journal}
  {Living Reviews in Relativity}\ }\textbf {\bibinfo {volume} {23}} (\bibinfo
  {year} {2020}{\natexlab{b}}),\ 10.1007/s41114-020-00026-9}\BibitemShut
  {NoStop}%
\bibitem [{\citenamefont {{Buchner, J.}}\ \emph {et~al.}(2014)\citenamefont
  {{Buchner, J.}}, \citenamefont {{Georgakakis, A.}}, \citenamefont {{Nandra,
  K.}}, \citenamefont {{Hsu, L.}}, \citenamefont {{Rangel, C.}}, \citenamefont
  {{Brightman, M.}}, \citenamefont {{Merloni, A.}}, \citenamefont {{Salvato,
  M.}}, \citenamefont {{Donley, J.}},\ and\ \citenamefont {{Kocevski,
  D.}}}]{refId0}%
  \BibitemOpen
  \bibfield  {author} {\bibinfo {author} {\bibnamefont {{Buchner, J.}}},
  \bibinfo {author} {\bibnamefont {{Georgakakis, A.}}}, \bibinfo {author}
  {\bibnamefont {{Nandra, K.}}}, \bibinfo {author} {\bibnamefont {{Hsu, L.}}},
  \bibinfo {author} {\bibnamefont {{Rangel, C.}}}, \bibinfo {author}
  {\bibnamefont {{Brightman, M.}}}, \bibinfo {author} {\bibnamefont {{Merloni,
  A.}}}, \bibinfo {author} {\bibnamefont {{Salvato, M.}}}, \bibinfo {author}
  {\bibnamefont {{Donley, J.}}}, \ and\ \bibinfo {author} {\bibnamefont
  {{Kocevski, D.}}},\ }\href {\doibase 10.1051/0004-6361/201322971} {\bibfield
  {journal} {\bibinfo  {journal} {A\&A}\ }\textbf {\bibinfo {volume} {564}},\
  \bibinfo {pages} {A125} (\bibinfo {year} {2014})}\BibitemShut {NoStop}%
\bibitem [{\citenamefont {Feroz}\ and\ \citenamefont
  {Hobson}(2008)}]{10.1111/j.1365-2966.2007.12353.x}%
  \BibitemOpen
  \bibfield  {author} {\bibinfo {author} {\bibfnamefont {F.}~\bibnamefont
  {Feroz}}\ and\ \bibinfo {author} {\bibfnamefont {M.~P.}\ \bibnamefont
  {Hobson}},\ }\href {\doibase 10.1111/j.1365-2966.2007.12353.x} {\bibfield
  {journal} {\bibinfo  {journal} {Monthly Notices of the Royal Astronomical
  Society}\ }\textbf {\bibinfo {volume} {384}},\ \bibinfo {pages} {449}
  (\bibinfo {year} {2008})},\ \Eprint
  {http://arxiv.org/abs/https://academic.oup.com/mnras/article-pdf/384/2/449/3378518/mnras0384-0449.pdf}
  {https://academic.oup.com/mnras/article-pdf/384/2/449/3378518/mnras0384-0449.pdf}
  \BibitemShut {NoStop}%
\bibitem [{\citenamefont {Feroz}\ \emph {et~al.}(2009)\citenamefont {Feroz},
  \citenamefont {Hobson},\ and\ \citenamefont
  {Bridges}}]{10.1111/j.1365-2966.2009.14548.x}%
  \BibitemOpen
  \bibfield  {author} {\bibinfo {author} {\bibfnamefont {F.}~\bibnamefont
  {Feroz}}, \bibinfo {author} {\bibfnamefont {M.~P.}\ \bibnamefont {Hobson}}, \
  and\ \bibinfo {author} {\bibfnamefont {M.}~\bibnamefont {Bridges}},\ }\href
  {\doibase 10.1111/j.1365-2966.2009.14548.x} {\bibfield  {journal} {\bibinfo
  {journal} {Monthly Notices of the Royal Astronomical Society}\ }\textbf
  {\bibinfo {volume} {398}},\ \bibinfo {pages} {1601} (\bibinfo {year}
  {2009})},\ \Eprint
  {http://arxiv.org/abs/https://academic.oup.com/mnras/article-pdf/398/4/1601/3039078/mnras0398-1601.pdf}
  {https://academic.oup.com/mnras/article-pdf/398/4/1601/3039078/mnras0398-1601.pdf}
  \BibitemShut {NoStop}%
\bibitem [{\citenamefont {Skilling}(2006)}]{10.1214/06-BA127}%
  \BibitemOpen
  \bibfield  {author} {\bibinfo {author} {\bibfnamefont {J.}~\bibnamefont
  {Skilling}},\ }\href {\doibase 10.1214/06-BA127} {\bibfield  {journal}
  {\bibinfo  {journal} {Bayesian Analysis}\ }\textbf {\bibinfo {volume} {1}},\
  \bibinfo {pages} {833 } (\bibinfo {year} {2006})}\BibitemShut {NoStop}%
\bibitem [{\citenamefont {Skilling}(2004)}]{doi:10.1063/1.1835238}%
  \BibitemOpen
  \bibfield  {author} {\bibinfo {author} {\bibfnamefont {J.}~\bibnamefont
  {Skilling}},\ }\href {\doibase 10.1063/1.1835238} {\bibfield  {journal}
  {\bibinfo  {journal} {AIP Conference Proceedings}\ }\textbf {\bibinfo
  {volume} {735}},\ \bibinfo {pages} {395} (\bibinfo {year} {2004})},\ \Eprint
  {http://arxiv.org/abs/https://aip.scitation.org/doi/pdf/10.1063/1.1835238}
  {https://aip.scitation.org/doi/pdf/10.1063/1.1835238} \BibitemShut {NoStop}%
\bibitem [{\citenamefont {Abbott}\ \emph
  {et~al.}(2019{\natexlab{b}})\citenamefont {Abbott} \emph
  {et~al.}}]{O2Isotropic}%
  \BibitemOpen
  \bibfield  {author} {\bibinfo {author} {\bibfnamefont {B.~P.}\ \bibnamefont
  {Abbott}} \emph {et~al.} (\bibinfo {collaboration} {LIGO Scientific and Virgo
  Collaboration}),\ }\href {\doibase 10.1103/PhysRevD.100.061101} {\bibfield
  {journal} {\bibinfo  {journal} {Phys. Rev. D}\ }\textbf {\bibinfo {volume}
  {100}},\ \bibinfo {pages} {061101} (\bibinfo {year}
  {2019}{\natexlab{b}})}\BibitemShut {NoStop}%
\bibitem [{\citenamefont {Abbott}\ and\ \citenamefont
  {et~al.}(2019)}]{publicO2}%
  \BibitemOpen
  \bibfield  {author} {\bibinfo {author} {\bibfnamefont {B.~P.}\ \bibnamefont
  {Abbott}}\ and\ \bibinfo {author} {\bibnamefont {et~al.}} (\bibinfo
  {collaboration} {LIGO Scientific Collaboration and Virgo Collaboration}),\
  }\href {https://dcc.ligo.org/LIGO-T1900058/public} {\emph {\bibinfo {title}
  {{Data for A search for the isotropic stochastic background using data from
  Advanced LIGO's second observing run}}}},\ \bibinfo {type} {Tech. Rep.}\
  \bibinfo {number} {LIGO-T1900058-v3}\ (\bibinfo {year} {2019})\BibitemShut
  {NoStop}%
\end{thebibliography}%


%

\end{document}